\documentclass[a4paper,10pt,aps,pre,longbibliography,twocolumn,amsmath,amssymb,nofootinbib,superscriptaddress]{revtex4-2}

\usepackage{xcolor,hyperref}
\hypersetup{
   colorlinks,
   linkcolor={blue!50!black},
   citecolor={blue!50!black},
   urlcolor={blue!80!black}
}

\usepackage{epsfig, amsmath}
\usepackage{bm}
\usepackage{soul,ulem}
\usepackage{hyperref}
\usepackage{footmisc}
\usepackage{wrapfig}

\DeclareMathAlphabet{\mathitbf}{OML}{cmm}{b}{it}

\newcommand{\zerovector}{\bm{0}}

\newcommand{\tripleCdot}{:\!\cdot\,}

\newcommand{\xv}{\mathitbf x}
\newcommand{\uv}{\mathitbf u}
\newcommand{\zv}{\mathitbf z}
\newcommand{\fv}{\mathitbf f}
\newcommand{\piv}{\bm{\pi}}

\newcommand{\calBold}[1]{\mbox{\boldmath${\cal #1}$}}

\newcommand{\dbar}{{\,\mathchar'26\mkern-12mu d}}

\begin{document}

\title{Detecting low-energy quasilocalized excitations in computer glasses}

\author{David Richard}
\email{david.richard@univ-grenoble-alpes.fr}
\affiliation{Univ.~Grenoble Alpes, CNRS, LIPhy, 38000 Grenoble, France}
\author{Geert Kapteijns}
\affiliation{Institute for Theoretical Physics, University of Amsterdam, Science Park 904, 1098 XH Amsterdam, The Netherlands}
\author{Edan Lerner}
\affiliation{Institute for Theoretical Physics, University of Amsterdam, Science Park 904, 1098 XH Amsterdam, The Netherlands}

\begin{abstract}
Soft, quasilocalized excitations (QLEs) are known to generically emerge in a broad class of disordered solids, and to govern many facets of the physics of glasses, from wave attenuation to plastic instabilities. In view of this key role of QLEs, shedding light upon several open questions in glass physics depends on the availability of computational tools that allow to study QLEs' statistical mechanics. The latter is a formidable task since harmonic analyses are typically contaminated by hybridizations of QLEs with phononic excitations at low frequencies, obscuring a clear picture of QLEs' abundance, typical frequencies and other important micromechanical properties. Here we present an efficient algorithm to detect the \emph{field} of quasilocalized excitations in structural computer glasses. The algorithm introduced takes a computer-glass sample as input, and outputs a library of QLEs embedded in that sample. We demonstrate the power of the new algorithm by reporting the spectrum of glassy excitations in two-dimensional computer glasses featuring a huge range of mechanical stability, which is inaccessible using conventional harmonic analyses due to phonon-hybridizations. Future applications are finally discussed.
\end{abstract}

\maketitle

\section{Introduction}
\label{sec:introdution}

Understanding the manners in which amorphous solids deform when subjected to mechanical stresses is a long-standing challenge in condensed matter physics~\cite{argon1976mechanism,spaepen1977microscopic}. Key to achieving progress in solving this problem relies on the development of new, efficient computational tools that are able to characterize amorphous structures at the microscopic level --- that are believed to govern glasses' mechanical response --- in a useful manner~\cite{richard2020predicting}. Many numerical developments have been put forward in this context, including novel bond-order-parameters~\cite{coslovich2007understanding,malins2013identification,tong2018revealing}, machine-learning based tools~\cite{ronhovde2011detecting,cubuk2015identifying,boattini2020autonomously,paret2020assessing,bapst2020unveiling,fan2021predicting,jung2022predicting}, approaches based on the potential energy landscape (PEL)~\cite{tanguy2010vibrational,manning2011vibrational}, and others~\cite{tsamados2009local,patinet2016connecting,xu2018predicting,schwartzman2019anisotropic}.

Recently it has been shown that glassy defects generically take the form of soft quasilocalized excitations (QLEs) that are composed of a core of few tens of particles decorated by a long-range algebraic Eshelby-like decay \cite{kapteijns2018universal}, see some visual representations in Fig.~\ref{fig:cost} below. In the low-frequency limit $\omega\!\to\!0$, QLEs are known to abide by a universal nonphononic density of states $D(\omega)\!=\!A_{\rm g}\omega^4$, where $A_{\rm g}$ has dimensions of [frequency]$^{-5}$ and controls the abundance of sofg glassy defects in the system~\cite{lerner2016statistics,kapteijns2018universal,wang2019low,rainone2020pinching,richard2020universality}. One of the main difficulties in studying these excitations within the harmonic approximation of the potential energy is that they hybridize with the overwhelming population of low-frequency phonons~\cite{phonon_widths_2018} -- present in any solid featuring a translationally invariant Hamiltonian. Due to said hybridizations, harmonic analyses do not allow to systematically study QLE's individual dynamics, nor how QLEs interact with each other.

To circumvent the aforementioned hybridization issues, novel computational frameworks have been put forward, which offer various micromechanical definitions of soft QLEs~\cite{gartner2016nonlinear,gartner2016nonlinear2,kapteijns2020nonlinear,richard2021simple} by incorporating anharmonic properties of the PEL. Within these nonlinear frameworks, a cost function associated with an arbitrary displacement field (referred to simply as a `mode') is constructed; these cost functions assume local minima at modes that simultaneously minimize the mode's energy, while maximizing its spatial localization. As demonstrated in~\cite{gartner2016nonlinear2,kapteijns2020nonlinear,richard2021simple}, these frameworks allow to filter-out the phononic background from hybridized phonon-QLEs harmonic modes. Furthermore, it has been demonstrated that the mechanics and spatial structure of quasilocalized excitations emerging from these nonlinear frameworks converge to those of harmonic excitations when hybridizations are absent~\cite{gartner2016nonlinear2,kapteijns2020nonlinear}, establishing the validity and usefulness of the nonlinear frameworks.

While these nonlinear-excitation-frameworks were helpful in establishing several results and insights~\cite{cge_paper,kapteijns2020nonlinear,JCP_Perspective}, they still do not yet allow for the exhaustive detection and extraction of the entire population of QLEs from a given computer-glass sample. In this work we address this issue and present an algorithm that builds on the same aforementioned nonlinear-excitation-frameworks; it takes as input a computer-glass sample, and outputs a library of the (nonlinear) QLEs embedded in that glass. We demonstrate the usefulness of our new algorithm by applying it to study the effect of thermal annealing on the abundance of QLEs in model glasses, and discuss further research directions that the new algorithm may open up.

This paper is structured as follows; we first provide the reader with a brief theoretical background of the nonlinear excitations framework, followed by a detailed presentation of our newly developed QLE-detection tools in Sec.~\ref{sec:algo}. In Sec.~\ref{sec:validation}, we investigate the impact of some of the input parameters of our algorithm and of the choice of the particular non-linear cost function employed on the extracted nonphononic spectrum of a model glass, and compare it directly with the harmonic vibrational density of states (vDOS) in three-dimensional systems. In Sec.~\ref{sec:result}, we employ the presented algorithm to examine the effect of thermal annealing on the statistics of QLEs in both two-dimensional (2D) and three-dimensional (3D) computer glasses. Finally, in Sec.~\ref{sec:disc} we discuss  the perspectives of our new algorithm to study viscous dynamics of supercooled liquids, in addition to the mechanical response of amorphous solids.

\section{Theoretical and numerical framework}
\label{sec:algo}
In this section, we spell out important theoretical background, which forms the basis of our QLE-detection algorithm. We also describe key features and caveats of our detection algorithm.

\subsection{Cost functions}

\begin{figure}[b!]
  \includegraphics[width = 0.5\textwidth]{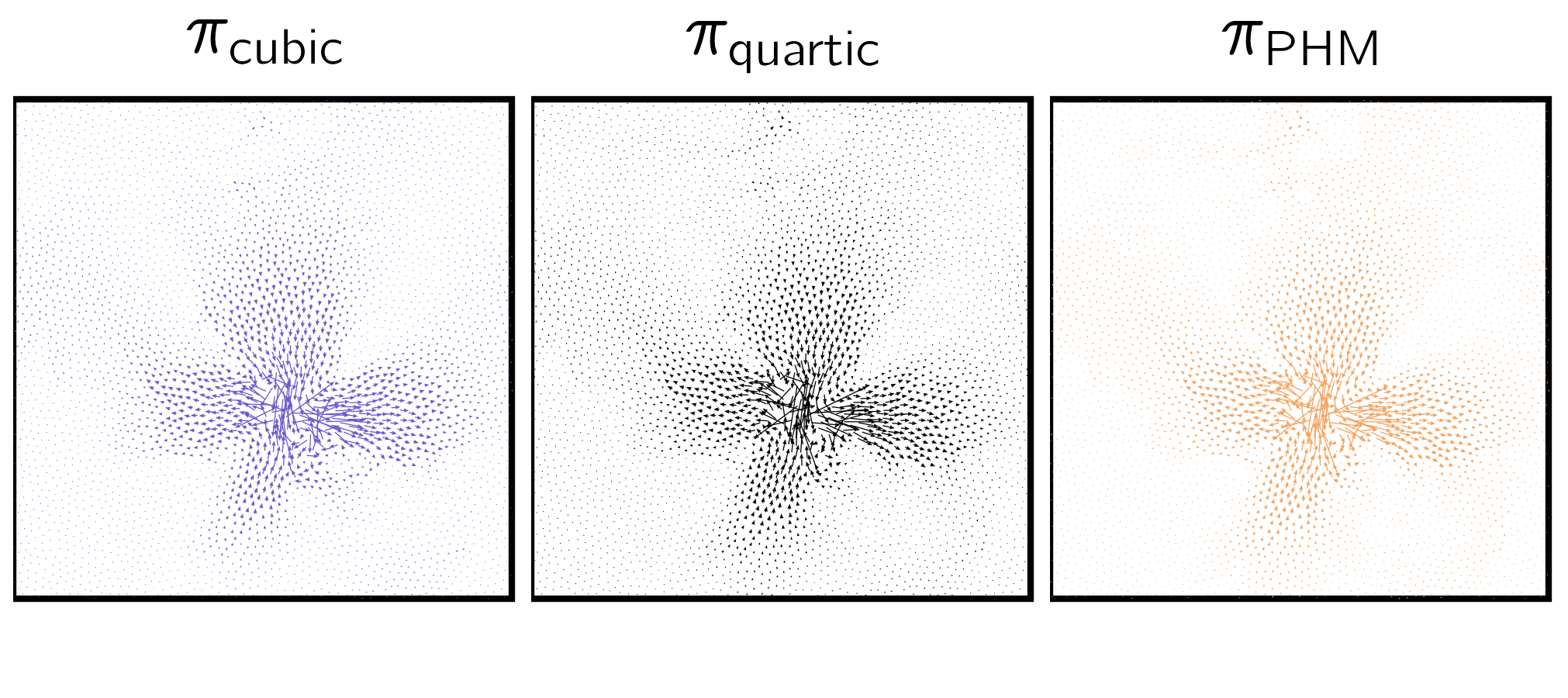}
  \caption{Different realizations of the same mode extracted from the cubic (a), quartic (b), and PHM (c) cost function. Mode energies are $\kappa_{\rm cubic}\!=\!0.428$, $\kappa_{\rm quartic}\!=\!0.347$, and $\kappa_{\rm PHM}\!=\!0.357$, respectively.}
  \label{fig:cost}
\end{figure}

We consider a system composed of $N$ particles in $\dbar$ spatial dimensions. As shown in Ref.~\cite{kapteijns2018universal} and visualized in Fig.~\ref{fig:cost}, soft spots in a glass can be realized as $N\dbar$-dimensional displacement fields $\zv$ that are composed of a localized core of a few tens of particles, decorated by a long range field that decays as $r^{-(\dbar-1)}$ at distance $r$ away from the core. Various nonlinear frameworks introduced in the past~\cite{gartner2016nonlinear,gartner2016nonlinear2,kapteijns2020nonlinear,richard2021simple} that enable the extraction of such excitations are all based on finding displacement fields $\piv$ at which a high dimensional cost function ${\cal C}(\zv)$ assumes local minima, namely
\begin{equation}\label{eq:eq1}
\frac{\partial {\cal C}}{\partial \zv}\bigg|_{\piv} = \zerovector\,.
\end{equation}
Here, ${\cal C}(\zv)$ is constructed in order to penalize both high energy \emph{and} delocalized (spatially extended) displacement fields. As such, plane waves --- which are inherently present in the harmonic approximation of any solid, and are generically spatially extended --- are suppressed, and do not form solutions to Eq.~(\ref{eq:eq1}). As presented in detail in Refs.~\cite{gartner2016nonlinear,gartner2016nonlinear2,kapteijns2020nonlinear,richard2021simple}, there are different ways to construct cost functions ${\cal C}(\zv)$ that can be practical~\cite{richard2021simple} and/or physically motivated~\cite{gartner2016nonlinear}. We next review the three cost functions ones used throughout this work, namely the cubic~\cite{gartner2016nonlinear}, quartic~\cite{gartner2016nonlinear2}, and pseudo-harmonic cost functions~\cite{richard2021simple}, distinguished in our notation by the subscript of ${\cal C}(\zv)$.

In the cubic cost-function framework, ${\cal C}_{\rm cubic}(\zv)$ is the `barrier function' that follows from a third order expansion of the energy with respect to displacements; it reads
\begin{equation}\label{cost_function}
{\cal C}_{\rm cubic}(\zv) = \frac{\kappa^3}{\tau^2}\,,
\end{equation}
where $\kappa(\zv)\!\equiv\!\calBold{H}:\zv\zv$ is the mode energy, and $\tau(\zv)\!\equiv\!\calBold{T}\!\tripleCdot\!\zv\zv\zv$ is the mode asymmetry, where $\calBold{T}\!\equiv\!\partial^3U/\partial\xv\partial\xv\partial\xv$ is the tensor of third-order derivatives of the energy with respect to particle coordinates. Here and in what follows, single, double, triple and quadruple contractions over Cartesian components and particle indices are denoted by $\cdot,:,\tripleCdot$ and $::$, respectively.

The quartic cost function follows from an analogy with ${\cal C}_{\rm cubic}$; here the third order contraction $\tau$ is replaced with the fourth  order contraction $\chi(\zv)\!\equiv\!\calBold{M}\!::\!\zv\zv\zv\zv$, where $\calBold{M}\!=\!\partial^4U/\partial\xv\partial\xv\partial\xv\partial\xv$ is the fourth order derivative of the energy with respect to particle coordinates, giving~\cite{gartner2016nonlinear2}
\begin{equation}\label{cost_function}
{\cal C}_{\rm quartic}(\zv) = \frac{\kappa^4}{\chi^2}.
\end{equation}
We note that $\chi$ has been shown to be inversely proportional to the mode participation ratio $e\!\equiv\!\big(N\sum_i (\zv_i\cdot\zv_i)^2\big)^{-1}$~\cite{gartner2016nonlinear2}, hence the denominator of ${\cal C}_{\rm quartic}$ will promote localization upon minimization. A detailed description about how to take high order derivatives of the energy, and about how it is contracted with a given field, is provided in Ref.~\cite{richardmicro2022}, for systems employing pairwise potentials.

It is typically cumbersome to compute the contractions $\tau$ and $\chi$ in systems employing many body interactions, or even ill-defined in cases that the employed interaction potentials are not smooth enough, e.g. Hertzian spheres. To overcome these difficulties, we have recently introduced a framework for extracting QLEs that solely utilizes the harmonic approximation of the potential energy~\cite{richard2021simple}. The excitations that emerged from this framework were coined pseudo-harmonic modes since they only require access to the harmonic approximation of the energy. The associated cost function reads
\begin{equation}\label{cost_function}
{\cal C}_{\rm PHM}(\zv) = \frac{\kappa^2}{\sum\limits_{\mbox{\tiny $\langle i,\! j\rangle$}}\big(\zv_{ij}\cdot\zv_{ij}\big)^2}\,,
\end{equation}
where $i,j$ are particle indices, $\zv_{ij}\!\equiv\!\zv_j\!-\!\zv_i$, and the sum in Eq.~(\ref{cost_function}) runs over all pairs $\langle i,\! j\rangle$ of interacting particles. For systems with long ranged interactions, one can run the sum over close neighbors defined e.g.~by a Voronoi analysis. Spatially extended, low-frequency plane waves have inherently small displacement gradients, i.e.~small local $\zv_{ij}$, and thus increase the cost function in favor of QLEs with larger $\zv_{ij}$ components at the mode's core.

In Fig.~\ref{fig:cost}, we show the same realization of a soft mode as a solution of Eq.~(\ref{eq:eq1}) for the three different cost functions ${\cal C}_{\rm cubic}(\zv),{\cal C}_{\rm quartic}(\zv)$ and ${\cal C}_{\rm PHM}(\zv)$ described above. We find a similar mode geometry with a nearly perfect mode overlap, i.e. $\pi_{\rm cubic}\!\cdot\!\pi_{\rm quartic}\!\simeq\!0.974$, $\pi_{\rm cubic}\!\cdot\!\pi_{\rm PHM}\!\simeq\!0.965$, and $\pi_{\rm quartic}\!\cdot\! \pi_{\rm PHM}\!\simeq\!0.999$. Note that subtle but important differences exist between cubic and quartic/PHM modes as the former includes anharmonic information about the potential energy landscape. As a result, cubic modes provide a better estimation (compared to the other considered cost functions) of configuration-space directions that tend to move the system across saddle points, at the cost of a slightly higher mode energy~\cite{kapteijns2020nonlinear}. This point is further discussed in Sec.~\ref{sec:validation}. A more detailed comparison between the properties of modes extracted from the different cost functions can be found in Refs.~\cite{gartner2016nonlinear2,kapteijns2020nonlinear,richard2021simple}.

\subsection{Mapping dipole forces}

Since the aforementioned cost functions are nonlinear, finding local minima of these function requires one to provide an initial guess --- denoted in what follows as $\zv_0$ --- for the minimization of the employed cost function, see Ref.~\cite{gartner2016nonlinear} for a visual demonstration. Our algorithm is constructed to provide spatially distinct initial guesses $\zv_0$ in order to efficiently and exhaustively find local minima of the employed cost function ${\cal C}$.

\begin{figure}[h!]
  \includegraphics[width = 0.5\textwidth]{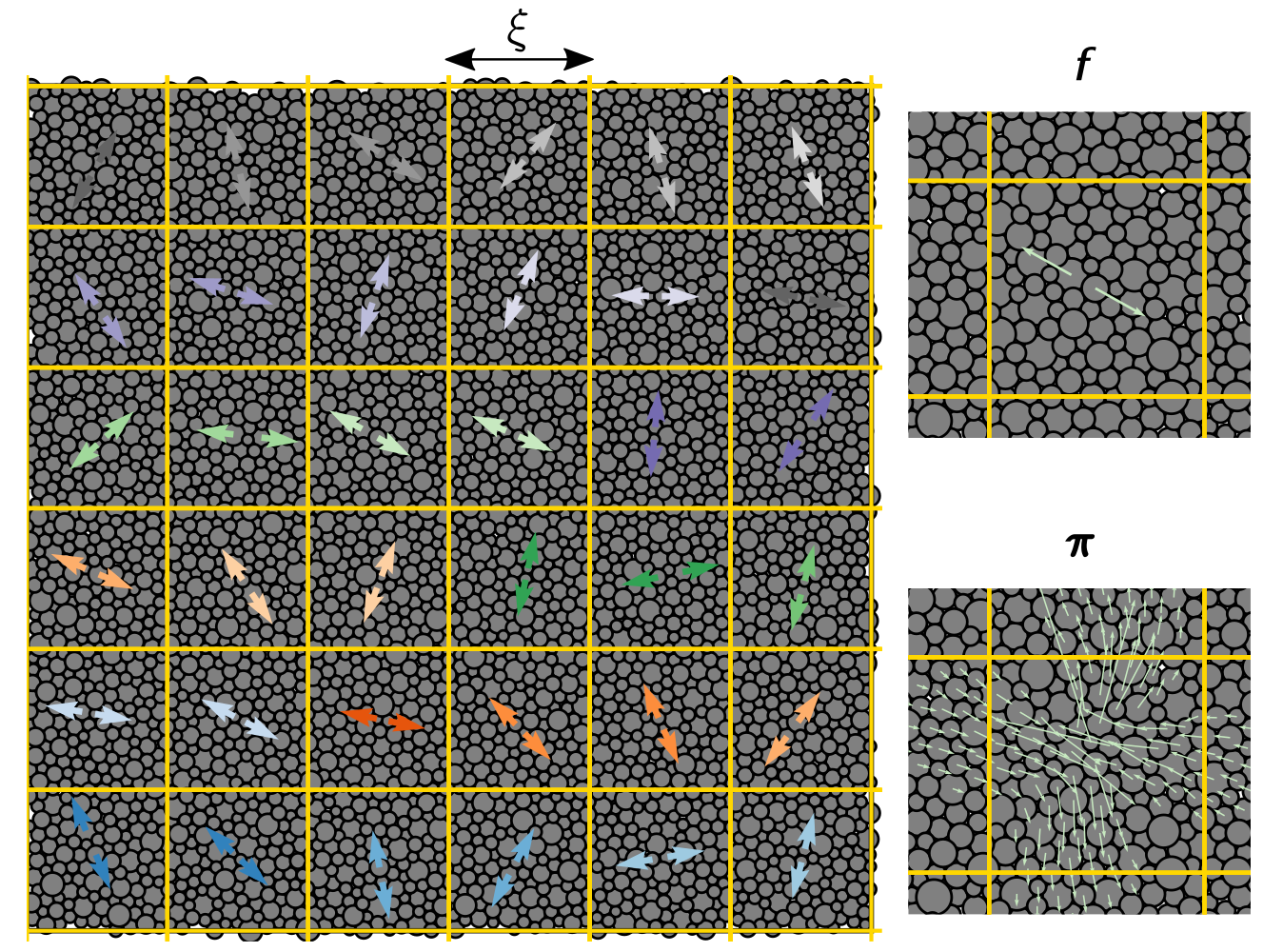}
  \caption{Sketch of \emph{stage 1} of our algorithm: (i) we probe glassy heterogeneities on a scale $\xi$, (ii) we pinch 2 particles with a force $\fv$ at the center of each block, and (iii) we map $\fv$ onto a solution $\piv$ of a given non-linear cost function ${\cal C}$. }
  \label{fig:algo1}
\end{figure}

To this aim, we first harvest local force dipoles, as illustrated in Fig.~\ref{fig:algo1}. We uniformly divide our system to blocks of linear size $\xi$; the latter is chosen to be consistent with the typical core size of QLEs (usually between 5-10 particle diameters \cite{rainone2020pinching}). Next, we consider the responses to local force dipoles $\fv$ acting on a pair of particles $\{ij\}$ centered in each block, see Fig.~\ref{fig:algo1}. These local force dipoles constitute excellent initial conditions $\zv_0$ for finding modes $\piv$ that are also located at the center of each cell (if such a mode exists), see example in Fig.~\ref{fig:algo1}. In parallel, we have harvested the displacement response $\uv$ to the same force dipole as obtained by solving the the linear equation $\calBold{H}\!\cdot\!\uv=\fv$. As shown in Ref.~\cite{rainone2020statistical}, displacements $\uv$ will be dominated by soft modes present in $\calBold{H}$ that project well onto the force $\fv$ and thus serve as potential initial guesses to find minima of ${\cal C}$. We have found similar results by directly starting our minimization from $\fv$, thus avoiding the extra cost of solving the aforementioned linear equation, and by such substantially reducing the computational complexity of our algorithm.

\begin{figure*}[t!]
  \includegraphics[width = 1\textwidth]{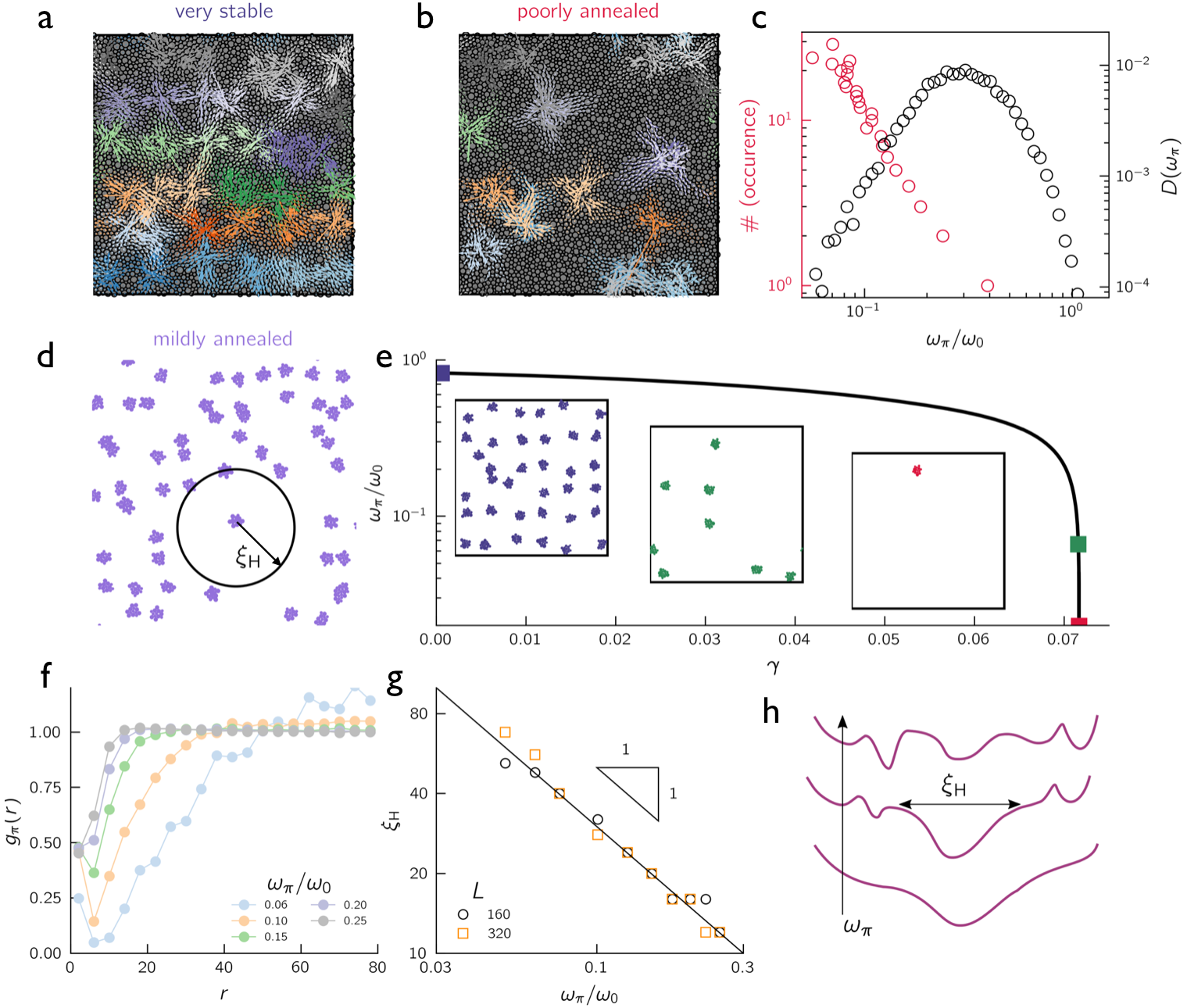}
  \caption{ First stage of our detection algorithm applied on a very stable (a) and poorly annealed (b) glass prepared at $T_{\rm p}=0.2$ and $0.7$, respectively. (c) Occurrence number (red) and density of states (black) as a function of the mode frequency $\omega$ for a glass prepared at $T_{\rm p}=0.7$. (d) Example of the halo effect in a large mildly annealed glass composed of $N=102400$. (e) Example of the halo effect when approaching a plastic instability. The black curve is the mode frequency as a function of the strain. Insets show our algorithm at different strains marked by colored squares. (f) Mode radial distribution function $g_\pi(r)$ for different mode $\pi$ with frequency $\omega_\pi$. (g) Halo radius $\xi_H$ as a function of $\omega_\pi$ for two different box lengths $L$. Here $\omega_0\!\equiv\!c_s/a_0$ with $c_s$ and $a_0$ denoting the shear-wave speed and typical interparticle distance, respectively. The black line indicates the scaling $\xi_H\sim 1/\omega_\pi$. (h) Sketch of the cost function landscape approaching the limit $\omega_\pi\!\to\!0$. All modes are extracted using ${\cal C}_{\rm PHM}$.}
  \label{fig:halo}
\end{figure*}

At this point, we have specified \emph{stage 1} of our algorithm, namely:
\begin{enumerate}
\item partition the system into blocks of linear size $\xi$.
\item Pinch a pair of particle with a force $\fv$ in the center of each block
\item Map the dipole force $\fv$ onto a mode $\piv$.
\end{enumerate}

\subsection{The ``halo" effect}
\label{sec:halo}

In Fig.~\ref{fig:halo} we present an example in which the first part of our QLE-detection algorithm described above is applied to two glass samples prepared by the SWAP Monte Carlo scheme~\cite{ninarello2017models}, which enables one to equilibrate liquids down to very low temperatures, i.e.~at very strong supercooling. Those equilibrium configurations are then instantaneously quenched to zero temperature, to form an ensemble of glasses labelled by the equilibrium parent temperature $T_{\rm p}$. With this scheme, we are able to build ensembles of glass samples featuring a very wide range of mechanical stability. The model's details and units employed can be found in Appendix~\ref{ap:protocol}. Here, we have considered two extreme cases of a very stable, ``cold" glass (Fig.~\ref{fig:halo}(a)) and a poorly annealed ``hot" glass (Fig.~\ref{fig:halo}(b)) prepared at $T_{\rm p}\!=\!0.2$ and $T_{\rm p}\!=\!0.7$, respectively. For the stable, low $T_{\rm p}$ glass, our framework is effective: each mapping from a dipole force leads to a \emph{distinct} solution $\piv$ associated with a frequency $\omega\!\equiv\!\sqrt{\calBold{H}\!:\!\piv\piv}$.

In contrast, we find in our poorly annealed sample that many dipole forces $\fv$ map (under the minimization of the ${\cal C}_{\rm PHM}$ cost function) to the \emph{same} solution $\piv$, resulting in ``empty" regions in the glass where there are presumably no QLEs. This effect, referred to as the ``halo effect" in what follows, is driven by the thermal-history induced changes in the properties of the cost functions ${\cal C}(\zv)$, and is quantified in Fig.~\ref{fig:bias}c, where we compare the (re-)occurrence number of modes vs.~their frequency $\omega_\pi$ with the density of states $D(\omega_\pi)$. Here, we have used an ensemble of 200 independent samples with $N\!=\!4096$ particles prepared at $T_{\rm p}=0.7$. Interestingly, we find that the lower the frequency, the higher the chance that a minimizations starting from dipole force $\fv$ located away from the mode's core -- will be mapped to the exact same mode. In other words, the direction-space \emph{volume} of the basins of the cost function ${\cal C}(\zv)$ --- that correspond to very low-frequency modes --- increases with decreasing mode frequency. Indeed, since the spectrum of quasilocalized excitations is gapless~\cite{JCP_Perspective}, we expect that a similar halo effect will be at play for glasses of any stability (i.e.~including very stable glasses), in the large system size limit $N\!\to\!\infty$. To illustrate this point we prepare a large mildly annealed 50-50 binary mixtures with $N=120400$ equilibated by conventional Molecular Dynamics~\cite{lerner2019mechanical}, and show in Fig.~\ref{fig:halo}d a close-up on one specific soft mode that reveals an approximated circular halo of radius $\xi_H$, in which no other solution of ${\cal C}_{\rm PHM}$ are found.

We also expect the emergence of the halo effect in mechanically driven solids, as the frequency of destabilizing modes on the brink of plastic instabilities can be arbitrary low, potentially giving rise to a large halo effect. This second situation is illustrated in Fig.~\ref{fig:halo}e, where we plot the frequency $\omega_\pi$ of a destabilizing mode as a function of the strain $\gamma$ upon a simple shear deformation. The insets of Fig.~\ref{fig:halo}e show the catalog of modes at different strain distance (indicated by colored squares in $\omega_\pi=f(\gamma)$) to a mechanical (plastic) instability. At the onset of the plastic instability, ${\cal C}_{\rm PHM}$ exhibits only one minimum, that corresponds to the critical mode (shown as a red blob of particles).

Next we want to quantitatively measure the relation between $\xi_H$ and the frequency $\omega_\pi$ of the underlying soft mode. Here, we compute the averaged radial distribution function $g_\pi(r)$ between a mode of frequency $\omega_\pi$ and all other solutions found with our algorithm, see Fig.~\ref{fig:halo}f. $g_\pi(r)$ is normalized by the ideal gas distribution,  with the number density computed from the total number of \emph{different} solutions $\piv$ found for a given glass realization. At large distances, $g_\pi$ goes to unity indicating that modes are randomly distributed with a uniform density. At small distances, we observe a gap that grows as $\omega_\pi$ decreases. We extract an estimate for $\xi_H$ at which $g_\pi(r)>0.9$ and plot it in Fig.~\ref{fig:halo}g against $\omega_\pi$. We find that the two are inversly related, namely $\xi_H\!\sim\!\omega_\pi^{-1}$, confirming that in the low-frequency limit the halo-effect will be system spanning. In Fig.~\ref{fig:halo}h, we sketch how the landscape of ${\cal C}$ changes in the limit $\omega_\pi\to 0$. The size $\xi_H$ of the metabassin diverges until ${\cal C}$ exhibits only a single minimum. The same behavior is observed for all three cost functions ${\cal C}(\zv)$ described in Sec.~\ref{sec:algo}.

\subsection{Biasing procedure}

To circumvent the halo effect described in the previous subsection -- namely to find QLEs in regions of the glass that seem ``empty" --, we propose to bias the employed cost function by systematically and artificially stiffening modes. In practice, this is done by connecting by a Hookean spring with stiffness $\kappa_b$ two particles residing at the center of the to-be-stiffened soft mode. In order to avoid over-stiffening the system, we choose $\kappa_b$ to be equal to the median of the interaction stiffness distribution. As a natural choice, we choose to stiffen the pair $\{ij\}$ corresponding to the largest relative longitudinal displacement between particles. The procedure is spelled out as follows; after mapping the force $\fv$ onto a mode $\piv$, we check whether or not this mode resides within its box of size $\xi$. If the resulting mode $\piv$ does not reside in the box associated with $\fv$, we add a Hookean spring to the core of $\piv$ and repeat the mapping of $\fv$ until a successful detection of a mode \emph{within} the box is made.

\begin{figure}[t!]
  \includegraphics[width = 0.5\textwidth]{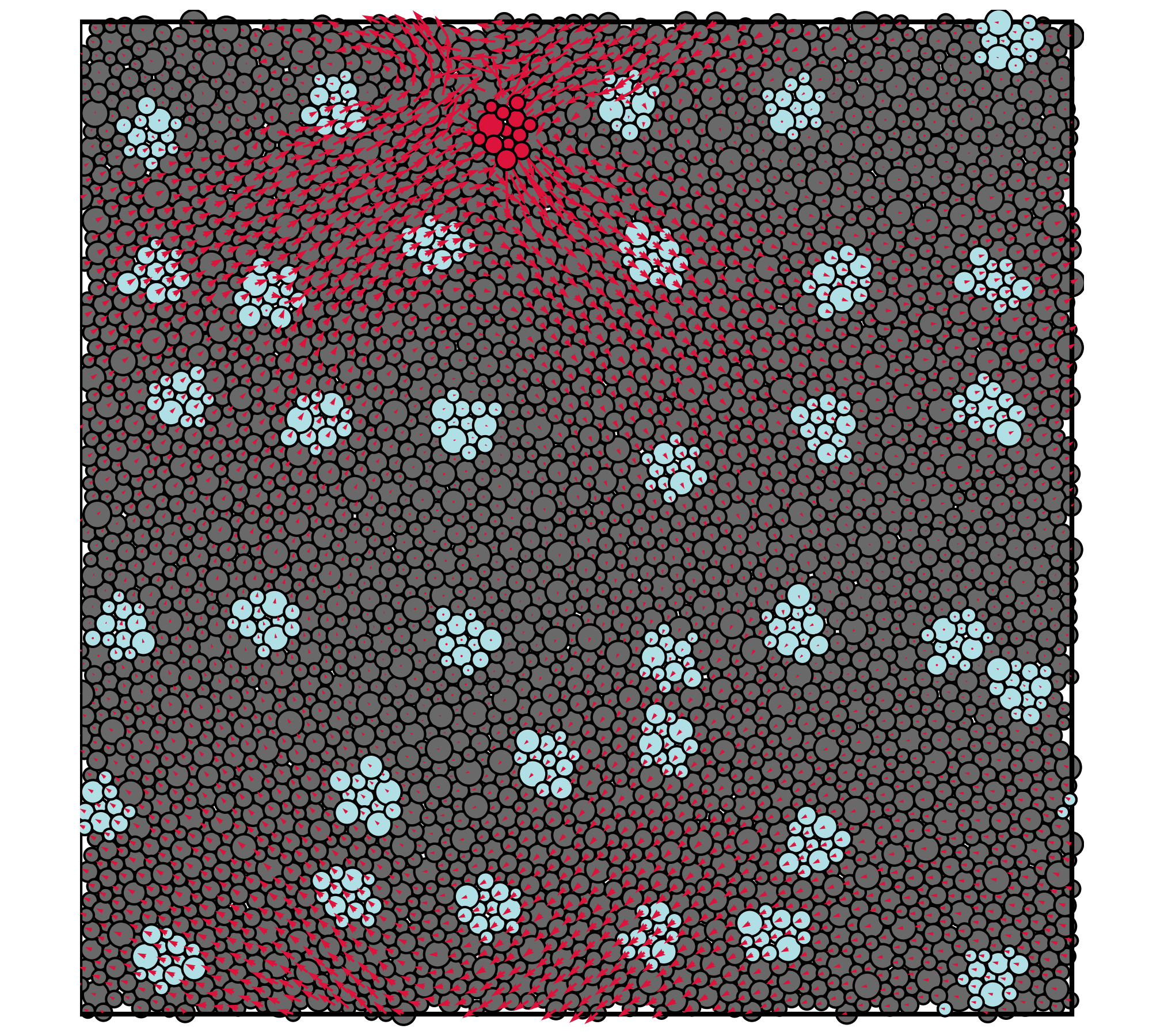}
  \caption{Example of our detection algorithm on a glassy configuration at the onset of a plastic instability with critical mode $\piv_c$ ($\omega_{\pi_c}/\omega_0\!\simeq\!0.02$) rendered in red. Each blob of particles colored in cyan correspond to the core of a distinct mode.}
  \label{fig:bias}
\end{figure}

As an extreme test case, we quasistatically shear a stable glass sample up to the first encountered plastic instability (up to strain $\gamma_{\rm c}\!\simeq\!0.07$), as previously shown in Fig.~\ref{fig:halo}(e). Recall that, upon approaching $\gamma_c$, we find that \emph{all} dipole forces $\fv$ are mapped onto a single mode $\piv$ -- the destabilizing mode $\piv_{\rm c}$ -- rendered in red in Fig.~\ref{fig:bias}. We now apply our biasing procedure and connect a Hookean spring of stiffness $\kappa_b$ between a pair of particles at the core of $\piv_{\rm c}$. Repeating stage 1 of the algorithm, we are able to recover excitations in the entire system, including ones in close spatial proximity to $\piv_{\rm c}$. We note, importantly, that each mode's frequency $\omega_\pi\equiv\!\sqrt{\calBold{H}\!:\!\piv\piv}$ is computed after removing all biasing (stiffening) springs.

We have now completed the description of stage 2 of our QLE-detection algorithm, namely:
\begin{itemize}
\item[4.] If a mapping of $\fv$ results in a mode $\piv$ residing \emph{outside} the block associated with $\fv$, we stiffen it with a spring of stiffness $\kappa_b$.
\item
[5.] Repeat step 4 until the detected mode $\piv$ resides within the block associated with $\fv$.
\end{itemize}
Combining stages 1 and 2 of the algorithm allows the extraction of the entire library of QLEs of a glass sample. The algorithm time $t_{\rm algo}$ needed to build a catalog for a fixed $\xi$ scales as $N^{2.4}$, as shown in Appendix~\ref{ap:complexity}. The scaling is explain by (i) the linear extensive nature of the catalog size for a fixed $\xi$, (ii) the extra linear complexity with $N$ of solving linear equations for a given mode $\piv$, and (iii) that the "halo" effect is more pronounced in larger systems, hence one needs to repeat step 4 more often. Importantly, we note that the QLEs catalog obtained with our algorithm does not depend on the sequence of blocks inspected and therefore can be trivially parallelized to investigate large system sizes.

\section{Validation of the algorithm}
\label{sec:validation}

As a validation benchmark, we directly compare the harmonic vDOS of computer glasses with the spectrum of pseudo-harmonic modes (PHMs) extracted using the QLE-detection algorithm spelled out above. Without compromising generality, we perform this comparison using mildly annealed 3D glasses with $N\!=\!2000$ particles (see Appendix~\ref{ap:protocol} for details). The motivation for this choice is to consider computer glasses featuring a relatively high abundance of QLEs, that can be easily measured using a conventional harmonic spectral analysis.

\begin{figure}[h!]
  \includegraphics[width = 0.5\textwidth]{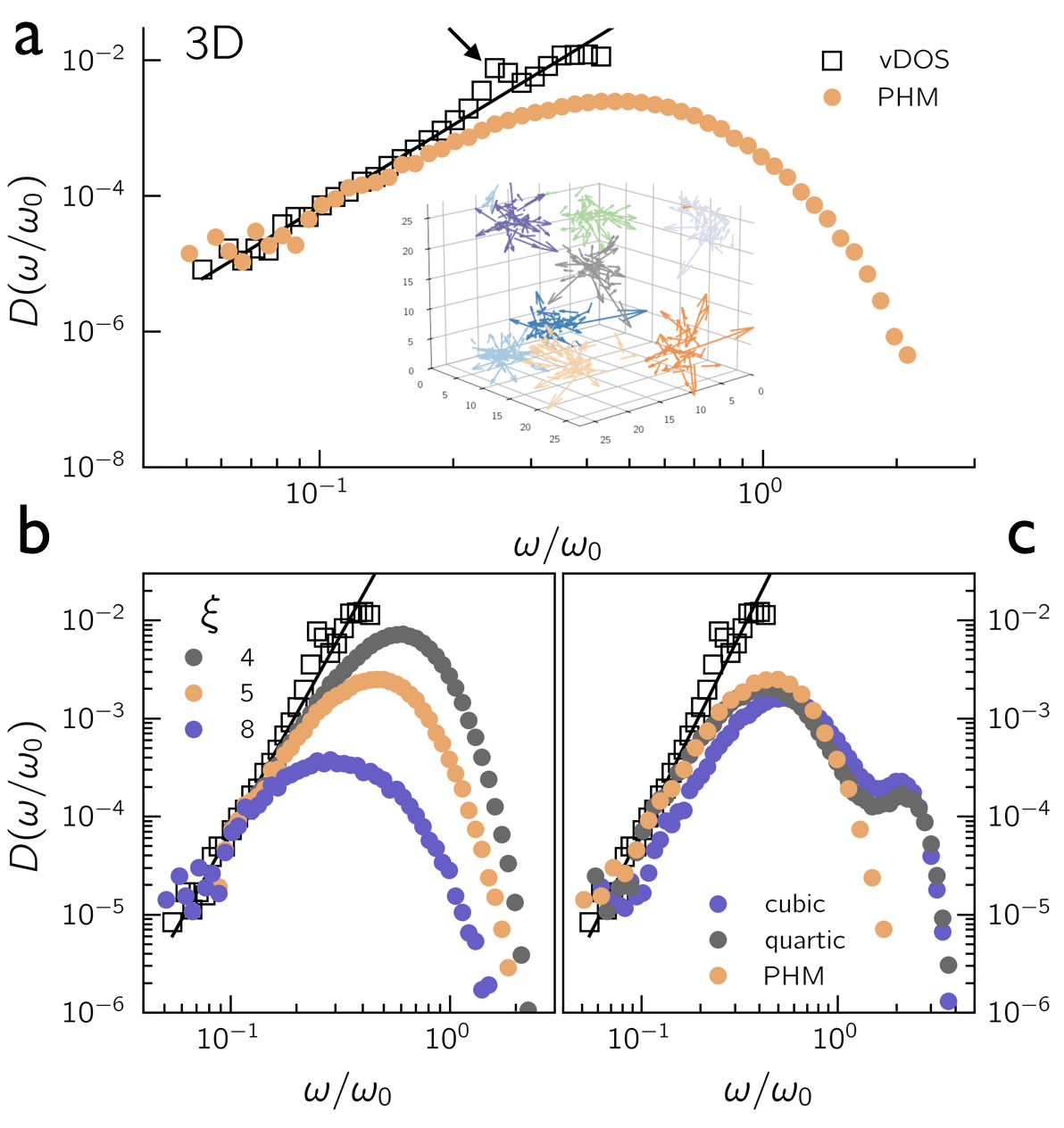}
  \caption{(a) Comparison between the harmonic and PHM vDOS in mildly annealed 3D glasses composed of $N\!=\!2000$ with $\xi\!=\!5$. The arrow indicates the first shear wave. The inset shows 8 modes detected in a sample composed of $N\!=\!16000$ with $\xi\!=\!10$. (b) Effect of partitioning-length $\xi$ on the PHM spectrum. (c) PHM spectrum fro different cost-function with $\xi\!=\!5$. The system is a 3D polydisperse glass prepared at $T_{\rm p}\!=\!0.6$.}
  \label{fig:benchmark}
\end{figure}

In Fig.~\ref{fig:benchmark}a, we report the PHM spectrum for the partitioning-length $\xi\!=\!5$ (expressed in terms of the typical interparticle distance $a_0$). The inset shows 8 modes extracted using our algorithm in a 3D sample. We find a good agreement between the harmonic and PHM vDOS below the first phonon (indicated by a small arrow): we find a gapless distribution at low-frequencies that displays the same $\sim\!\omega^4$ scaling as seen for the harmonic modes vDOS. At higher frequencies, the PHM vDOS exhibits a maximum before decreasing and vanish at a characteristic upper cutoff frequency.

We next vary $\xi$ from 4 to 8, see fig.~\ref{fig:benchmark}b. In a system of $N\!=\!2000$ this corresponds in extracting between 1 to 8 modes per sample. For $\xi\!=\!8$, we already recover a large fraction of modes that populate the low frequency tail of the harmonic vDOS. As $\xi$ decreases, we progressively pick up stiffer modes that extend well above the first shear wave. Decreasing further $\xi$ will thus only result in the detection of high-frequency excitations. 

Finally, we set $\xi=5$ and investigate the nonphononic spectrum obtained using different cost functions. In Fig.~\ref{fig:benchmark}c, we compare results obtained using the PHM-cost function with those obtained using the cubic and quartic cost functions, detailed in Sec.~\ref{sec:algo}. At low frequencies we find that PHMs and quartic modes have the lowest energy and produce the same low frequency tail as the conventional harmonic vDOS. 

In contrast, the excitations obtained using the cubic cost function (referred to as \emph{cubic modes}) are found to be slightly stiffer (as also pointed out in Ref.~\cite{gartner2016nonlinear2}).
This slightly higher energy of cubic modes is explained by the need of maximizing the mode asymmetry when minimizing ${\cal C}_{\rm cubic}$. However, cubic modes still exhibit the correct quartic scaling $D(\omega)\!\sim\!\omega^{4}$.

\begin{figure}[t!]
  \includegraphics[width = 0.5\textwidth]{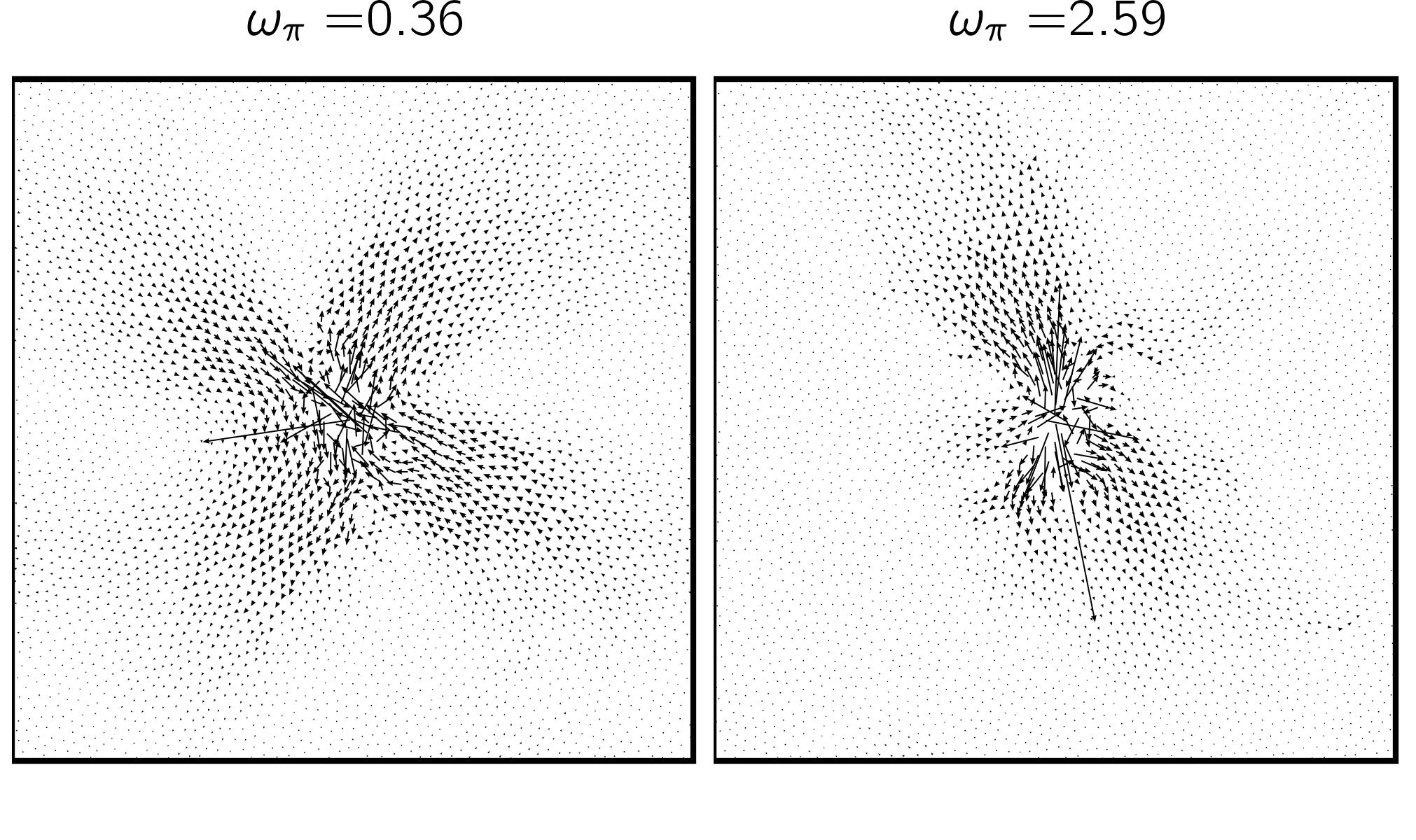}
  \caption{Low- (left) and high-frequency (right) cubic modes extracted in a mildly annealed 2D polydisperse glass prepared at $T_{\rm p}=0.6$. Modes have been shifted to the center of the simulation box for the sake of visibility.}
  \label{fig:doublepeak}
\end{figure}

At high frequencies, we find for the cubic and quartic modes a second peak in their respective specta. The same double peak distribution is observed in 2D glasses. Inspecting modes populating the 1st and 2nd peak, we find that stiff modes show a less quadrupolar anisotropy compared with their low-frequency counterpart, see fig.~\ref{fig:doublepeak}. We link this second population to the presence of large pairwise forces that increases the denominator of ${\cal C}_{\rm cubic}$ and ${\cal C}_{\rm quartic}$, the force magnitude being absent in ${\cal C}_{\rm PHM}$. The resulting mode geometry corresponds to two particles being pushed away from each other. We note that a similar behavior of the excitation distribution was observed when investigating the statistical mechanics of dipole responses~\cite{rainone2020statistical}.

\section{Applications}
\label{sec:result}

\subsection{Nonphononic spectrum of 2D structural glasses}

Recently it has been debated whether or not 2D computer solids feature the same $\sim\!\omega^4$ scaling of their nonphononic vDOS as 3D solids do~\cite{Grzegorz_prl_2021,Grzegorz_erratum_2022,wang2022scaling,lerner2022nonphononic,atsushi_pinning}.  One of the difficulties to investigate the QLE-statistics in 2D is that one needs to employ large-enough samples to avoid finite-size effects~\cite{lerner2020finite,lerner2022nonphononic}, typically of a few thousand of particles in size. The lateral box length $L$ for such system sizes is typically on the order $50\!-\!100$ particle diameters. The frequency of the first shear wave decreases as $\omega_{\rm ph}\sim L^{-1}$ and the onset of QLEs scales as $\sim L^{-2/5}$~\cite{lerner2016statistics}. As a consequence, most of the QLEs that populate the low-frequency, harmonic nonphononic spectrum are hybridized with plane waves. More details can be found in Ref.~\cite{phonon_widths_2018}. This limiting issue is even more pronounced for very stable glasses, since in those systems the population of QLEs is both stiffened and depleted~\cite{cge_paper,rainone2020pinching,rainone2020statistical}. So far, due to the aforementioned difficulties and in contrast to 3D solids, there have been no reports on how the nonphononic prefactor $A_{\rm g}$ varies as a function of the parent temperature $T_{\rm p}$ in 2D. Below we show that our QLE-detection algorithm is able to provide this information.

\begin{figure}[h!]
  \includegraphics[width = 0.5\textwidth]{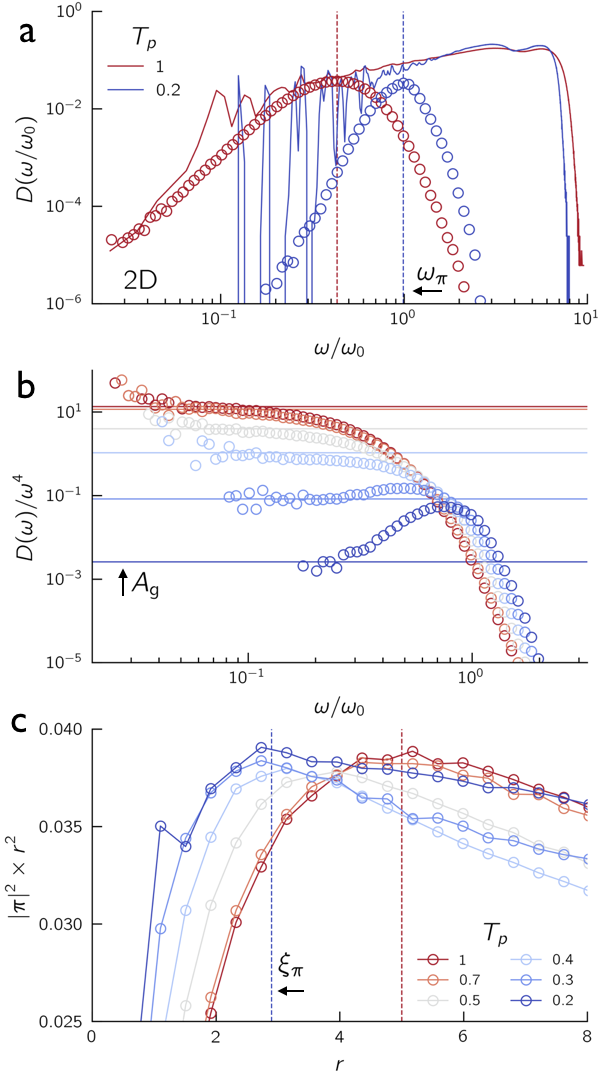}
  \caption{(a) Comparison between the full harmonic (solid lines) and PHM (empty circles) vDOS for 2D glasses quenched from parent temperatures $T_{\rm p}$ as indicated by the legend. PHM spectra are computed using $\xi=5$. The dashed vertical lines indicate the shift of the \emph{average} mode frequency $\omega_\pi$ between our most stable glassy configurations (blue) and hyper-quenched glasses (red). (b) Rescaled PHM vDOS for various parent temperature $T_{\rm p}$ with $\xi=5$. The horizontal lines indicate $A_{\rm g}$.  (c)  Rescaled square of the mode amplitude decay $|\pi|^2$ by $r^2$. Vertical dashed lines indicate the reduced core size $\xi_\pi$ when decreasing $T_{\rm p}$.}
  \label{fig:2dvdos}
\end{figure}

Additionally, thanks to the QLE-detection algorithm we are now able to also identify the upper frequency limit up to which QLEs exist, and observe its variability with changing glass stability. Our results are shown in Fig.~\ref{fig:2dvdos}; in Fig.~\ref{fig:2dvdos}a we compare the 2D harmonic vDOS with the PHM spectrum of a poorly annealed ($T_{\rm p}\!=\!1.0$) and very stable ($T_{\rm p}\!=\!0.2$) glasses (see Appendix~\ref{ap:protocol} for details about the model and its units). We observe a strong suppression of the prefactor of $\omega^4$ tail as well as a shift of the QLM spectrum towards larger frequencies.

In Fig.~\ref{fig:2dvdos}b, we show the complete PHM spectrum rescaled by $\omega^4$ for various $T_{\rm p}$. We find a plateau at low frequency confirming that the nonphononic statistics in 2D follows $D(\omega)=A_{\rm g}\omega^4$ for $\omega\to0$. We note that small deviations from the $\omega^4$ scaling are observed for hyper-quenched glasses ($T_p\!>\!0.5$), fully consistent with the known finite-size effects~\cite{lerner2020finite,lerner2022nonphononic}. In addition, one can read the values of $A_{\rm g}$ off the $\omega\!\to\!0$ plateau values of $D(\omega)/\omega^4$; we find a similar variability of $A_{\rm g}$ compared to 3D observations~\cite{rainone2020pinching,giannini2021bond} --- of over 3 decades ---  with a drop of about 3 orders of magnitude when decreasing $T_{\rm p}$.

Collecting from our catalog the lowest mode in each glassy configuration, we can compute the average of the square of the mode amplitude decay $|\pi|^2$ that populate the $\omega^4$ scaling. As shown in Ref.~\cite{rainone2020pinching}, one can rescale $|\pi|^2$ by its algebraic decay, $\sim r^{-2}$, and extract an estimate for the typical core size $\xi_\pi$ at which $r^2|\pi|^2$ shows a maximum, see Fig.~\ref{fig:2dvdos}c. We clearly observe a decrease of $\xi_\pi$ for very stable glasses. This result is in line with previous numerical studies for 3D amorphous solids~\cite{rainone2020pinching,giannini2021bond}.

\subsection{Robustness and caveats}

\begin{figure}[h!]
  \includegraphics[width = 0.5\textwidth]{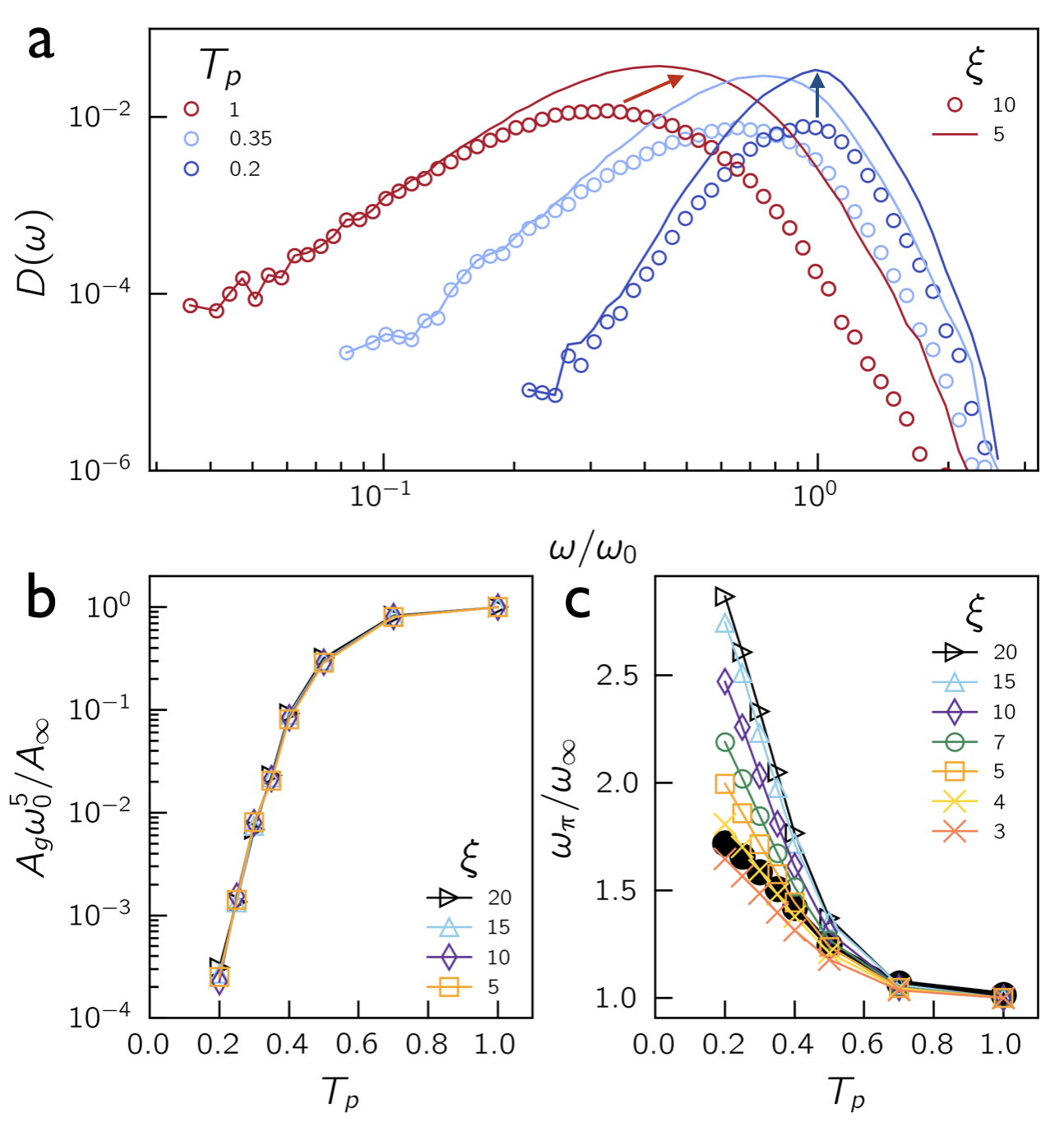}
  \caption{(a) Comparison between the PHM vDOS extracted with $\xi\!=\!10$ (empty circles) and $\xi\!=\!5$ (solid lines) for different parent temperatures. (b) The prefactors $A_{\rm g}$ plotted versus the parent temperature $T_{\rm p}$ for various $\xi$. (c) The normalized characteristic frequency $\omega_\pi/\omega_\infty$ of quasilocalized modes for various $\xi$. Filled black circles are the normalized characteristic frequency of dipole responses taken from Ref.~\cite{rainone2020statistical}.}
  \label{fig:robustness}
\end{figure}

We now discuss the robustness and caveats of our algorithm in regards the extraction of $A_{\rm g}$ and $\omega_\pi$ as a function of $T_{\rm p}$. As already pointed out in Fig.~\ref{fig:benchmark}(b), decreasing $\xi$ results in the detection of stiffer modes. In Fig.~\ref{fig:robustness}(a), we show the 2D PHM spectra for $\xi\!=\!5$ and $\xi\!=\!10$ measured in glassy samples ranging from poorly annealed ($T_{\rm p}\!=\!1.0$) to very stable ($T_p\!=\!0.2$). As observed in 3D glasses the low-frequency spectrum is unchanged upon decreasing $\xi$. As a consequence, the range of $A_{\rm g}$ versus $T_p$ does not vary with $\xi$, as shown in Fig.~\ref{fig:robustness}(b).

For stable glasses, we also observe that the typical frequency scale $\omega_\pi$ remains constant under variations of $\xi$. This is not the case for poorly annealed samples, where $\omega_\pi$ slowly increases with decreasing $\xi$, see arrows in Fig.~\ref{fig:robustness}a. As a result, the relative stiffening of $\omega_\pi/\omega_\infty$ with thermal annealing decreases with $\xi$, see Fig.~\ref{fig:robustness}c. Interestingly, we find that when $\xi$ approaches the typical QLM core size $\xi_\pi$ (about 3-5 particle diameters), $\omega_\pi/\omega_\infty$ exhibits the same range as previously found when extracting an energy scale from the dipole response statistics~\cite{rainone2020statistical}. We have found similar results in our 3D glassy solids.

\subsection{Effect of thermal annealing 2D vs. 3D}

In what follows, we define the typical mode frequency $\omega_\pi$ as the \emph{average} frequency of the full QLE spectrum. We are now in position to extract both the abundance of soft excitations through the dimensionless prefactor $A_{\rm g}$, the characteristic mode frequency $\omega_{\pi}$, and the core size $\xi_\pi$. Additional data for 3D glasses are provided in Appendix~\ref{ap:3dvdos}. Here, we propose a comparison of 2D and 3D solids as a function of glass stability controlled by $T_{\rm p}$. Note that this direct comparison is meaningful as both our 2D and 3D systems share the same crossover temperature $T_{\rm cross}\simeq 0.6$.

\begin{figure}[h!]
  \includegraphics[width = 0.5\textwidth]{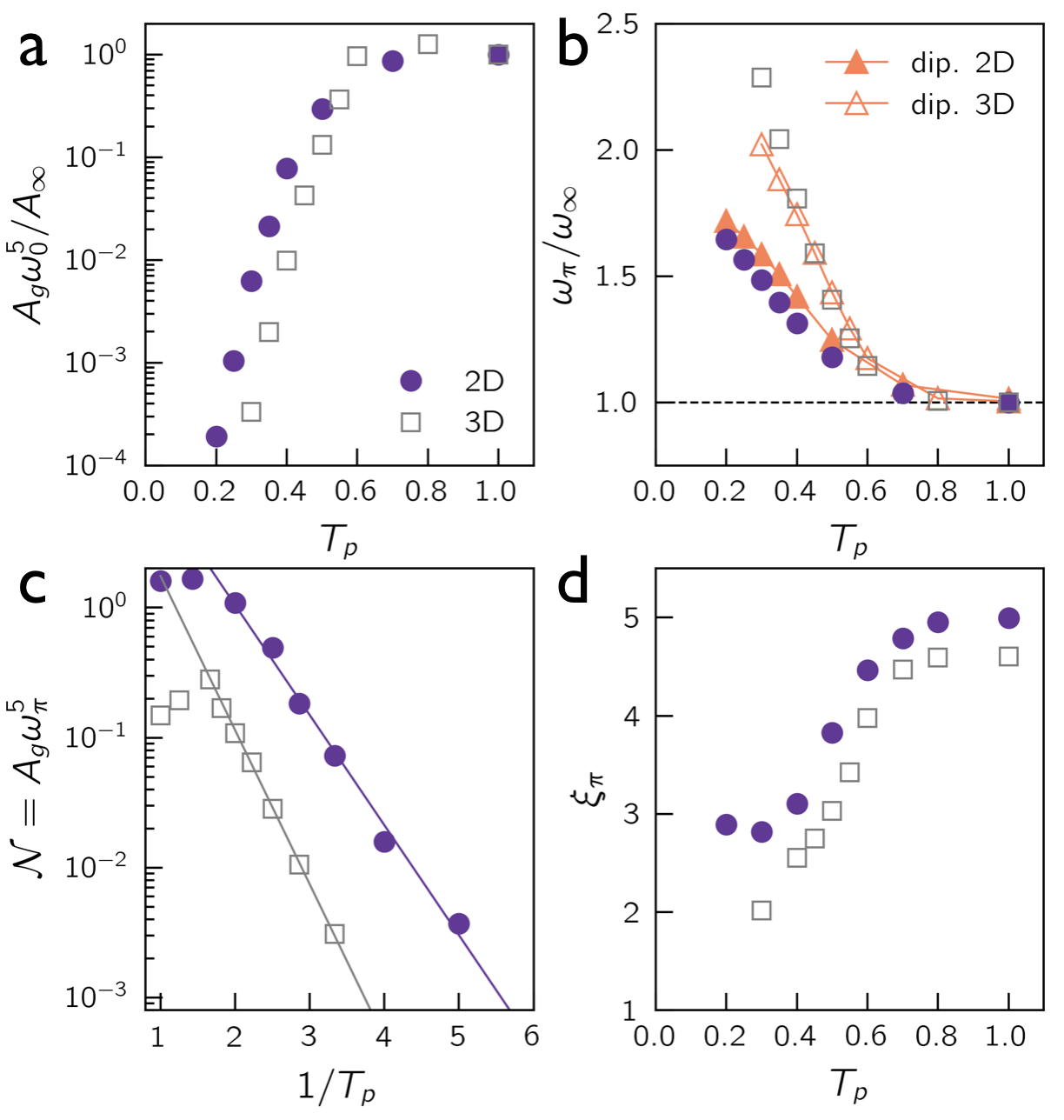}
  \caption{(a) The prefactors $A_{\rm g}$ plotted versus the parent temperature $T_{\rm p}$ for both 2D and 3D glasses with $\xi=5$. $A_{\rm g}$ is normalized by the high temperature value $A_\infty=A_{\rm g}(T_{\rm p}=1)$. (b) The normalized characteristic frequency $\omega_\pi/\omega_\infty$ of quasilocalized modes for $\xi=3$. The orange data represent the characteristic frequency scale of dipole responses, taken from Ref.~\cite{rainone2020statistical}. (c) The density $\mathcal{N}$ of QLEs, plotted as a function of $1/T_{\rm p}$. Solid lines are linear regressions. (d) Core size $\xi_\pi$ versus $T_{\rm p}$.}
  \label{fig:2d_vs_3d}
\end{figure}

In Fig.~\ref{fig:2d_vs_3d}a, we plot $A_{\rm g}$ as a function of $T_{\rm p}$ normalized by the high temperature plateau value $A_{\infty}$. As discussed previously, we find the same qualitative behavior in 2D and 3D. Below  $T_{\rm cross}$, $A_{\rm g}$ drops by more than 3 orders of magnitude between the high temperature plateau and the lowest temperature accessible by SWAP Monte Carlo, such as found in Ref.~\cite{rainone2020pinching}. Moreover, the suppression of modes is postponed to slightly lower temperature in 2D compared to 3D solids.

As discussed extensively in Refs.~\cite{cge_paper,rainone2020pinching}, $A_{\rm g}$ is of units of an inverse frequency to the fifth power. Accordingly, a drop in $A_{\rm g}$ as a function of $T_{\rm p}$ is not necessarily caused by a decrease in the number density of QLEs populating the $\omega^4$ scaling, but could be attributed to an overall mode stiffening. As such, we first need to quantify the characteristic frequency $\omega_\pi$ of QLEs. In Fig.~\ref{fig:2d_vs_3d}b, we plot the characteristic mode frequency $\omega_\pi$ normalized by the high parent temperature plateau $\omega_\infty$ as a function of $T_{\rm p}$. We find a stiffening of about a factor of two for both our 2D and 3D glasses. Consistent with the variability of $A_{\rm g}$, the stiffening is weaker for 2D solids (about 25-30\%). This result is consistent with what has been found using the statistics of dipole responses~\cite{rainone2020statistical}. One should, however, keep in mind that, strictly speaking, QLE's typical frequency $\omega_\pi$ in 2D features a logarithmic dependence on system size $L$~\cite{cge_paper,kapteijns2018universal,lerner2022nonphononic}. Here, we have kept $L$ fixed, and consider the relative variability. 

We now can compute the QLEs density as $\mathcal{N}\!=\!A_{\rm g}\omega_{\pi}^{5}$ and plot it against the inverse parent temperature $1/T_{\rm p}$, see Fig.~\ref{fig:2d_vs_3d}c. As seen for $A_{\rm g}$, $\cal{N}$ plateau at high $T_{\rm p}$, but shows a weaker decrease with a drop of about 2-3 orders of magnitude when decreasing $T_{\rm p}$. In both 2D and 3D, we observe an Arrhenius behavior where the QLEs depletion follows $\mathcal{N}\!\sim\!e^{-E_{\rm qlm}/k_BT_{\rm p}}$, with $E_{\rm qlm}$ a formation energy such as that found in Ref.~\cite{rainone2020pinching}.

Finally, we report in Fig.~\ref{fig:2d_vs_3d}d the core size $\xi_\pi$ as a function of $T_{\rm p}$. For hyperquenched glasses (high $T_{\rm p}$), the core size is about 5-6 particle diameters. This result is consistent with the typical coarse graining length scale used to quantitatively parametrized mesoscale elasto-plastic models from microscopic simulations~\cite{castellanos2021insights,liu2021elastoplastic}. We observe a factor two decrease of $\xi_\pi$ when lowering $T_{\rm p}$. This result is consistent with Ref.~\cite{rainone2020pinching}, where authors argued that the core size and the characteristic frequency scale are inversely proportional $\xi_\pi \sim 1/\omega_\pi$. Accordingly, we find a slightly weaker decrease of $\xi_\pi$ versus $T_{\rm p}$ in 2D compared with 3D glasses.

\section{Discussion and conclusion}
\label{sec:disc}

In this paper, we have presented a novel and efficient algorithm to detect the field of soft glassy excitations in structural glasses. Utilizing dipole forces as local probes, we have shown that one can construct the map of soft excitations in both 2D and 3D. We have highlighted that our detection algorithm suffers from a ``halo" effect (see Sect.~\ref{sec:halo}) created by the softest excitations present in the system: once a mode with vanishing frequency exists in a given system, the minimization of the cost function ${\cal C}(\zv)$ will only converge to the same low-frequency solution. In other words, extremely soft modes result in cost functions ${\cal C}(\zv)$ featuring a single basin, which corresponds to the soft mode. 

To circumvent this issue, we have put forward an easy biasing procedure where soft modes in the system are being stiffen by harmonic springs. This enables us to find soft excitations even for a system driven towards a plastic instability, during which the frequency of the destabilizing mode vanishes. We have benchmarked our algorithm with the harmonic spectrum of 3D glasses and have explored the influence of the algorithm parameters as well as the non-linear cost function used. We have demonstrated that one can recover the correct asymptotic low-frequency nonphononic tail.

Using our new algorithm, we have extracted for the first time the complete spectrum of localized soft excitations in 2D solids featuring a wide range of mechanical stability (generated using SWAP Monte Carlo). We have confirmed that the low-frequency limit of nonphononic excitations follows $D(\omega)=A_{\rm g}\omega^4$, supporting recent claims~\cite{lerner2022nonphononic,atsushi_pinning}. Finally, we have reported how the dimensionless prefactor $A_{\rm g}\omega_{0}^5$ and the characteristic mode frequency $\omega_\pi$ change with thermal annealing. The same qualitative behavior is seen between 2D and 3D solids. We find a drop in the number density of QLEs $\mathcal{N}\!=\!A_{\rm g}\omega_{\pi}^{5}$ of about 2-3 orders of magnitude, a mode stiffening $\omega_\pi$ of a factor two, and a factor two decrease of the core size $\xi_\pi$. This result confirms the previously established relation $\omega_\pi\!\sim\!c_s/\xi_\pi$ between the QLE's characteristic frequency and the glassy length scale. Moreover, a 25-30\% weaker stiffening is observed in 2D glasses compared with their 3D counterpart. One could be tempted to draw an analogy between a stronger elastic stiffening mechanism in 3D, and the possible occurrence of a finite-temperature random first-order transition, which is likely to be absent in 2D~\cite{guiselin2022statistical}.

This work opens new avenues to understand elastic heterogeneities in amorphous solids. In particular, one can use non-linear modes to access information on distributions of activation barriers, strain distances to instability~\cite{kapteijns2020nonlinear}, and tensorial information such that the softest shear direction~\cite{richardmicro2022}. With our method, one will be able to monitor how these distributions change upon aging or mechanical deformation.

Furthermore, compared to modes built solely on the Hessian, cubic modes have the advantage to offer better core direction to move towards the nearby saddle. Thus, one could harvest the field of cubic modes to efficiently explore the potential energy landscape of a glass and improve existing methods such as saddle point sampling~\cite{xu2018predicting}.

\acknowledgements
D.R. acknowledges support by the European Union’s Horizon 2020 research and innovation programme under the Marie Sklodowska-Curie grant agreement No. 101024057. Support from the NWO (Vidi grant no.~680-47-554/3259) is acknowledged.


\appendix

\section{Glass former}
\label{ap:protocol}

Results shown in the main text are for two types of glass formers, namely polydisperse soft spheres and a 50-50 binary mixture. In both model, particles interact via an inverse power law potential. A detailed description of these models is provided in Ref.~\cite{lerner2019mechanical}. Thanks to SWAP Monte Carlo (MC)~\cite{ninarello2017models}, we generate 2D and 3D polydisperse glasses with various degrees of stability. The later is controlled by the parent temperature $T_{\rm p}$ of the equilibrium states from which our glasses were instantaneously quenched. The binary mixture is quenched at a finite rate using conventional Molecular Dynamics. All simulations are performed in the NVT ensemble with number density $N/V\!=\!0.65$ (2D) and $0.58$ (3D). Finally, we quench our configurations to zero temperature via an energy minimization using the conjugate gradient algorithm. All quantities in this paper are reported in dimensionless microscopic units: lengths are rescaled by $a_0=(V/N)^{1/\dbar}$, frequencies by $\omega_0\!=\!\sqrt{\mu_\infty/\rho}/a_0$, where $\mu_\infty$ is the plateau shear modulus of the high-parent temperature glasses.

\section{Algorithm complexity}
\label{ap:complexity}

\begin{figure}[h!]
  \includegraphics[width = 0.5\textwidth]{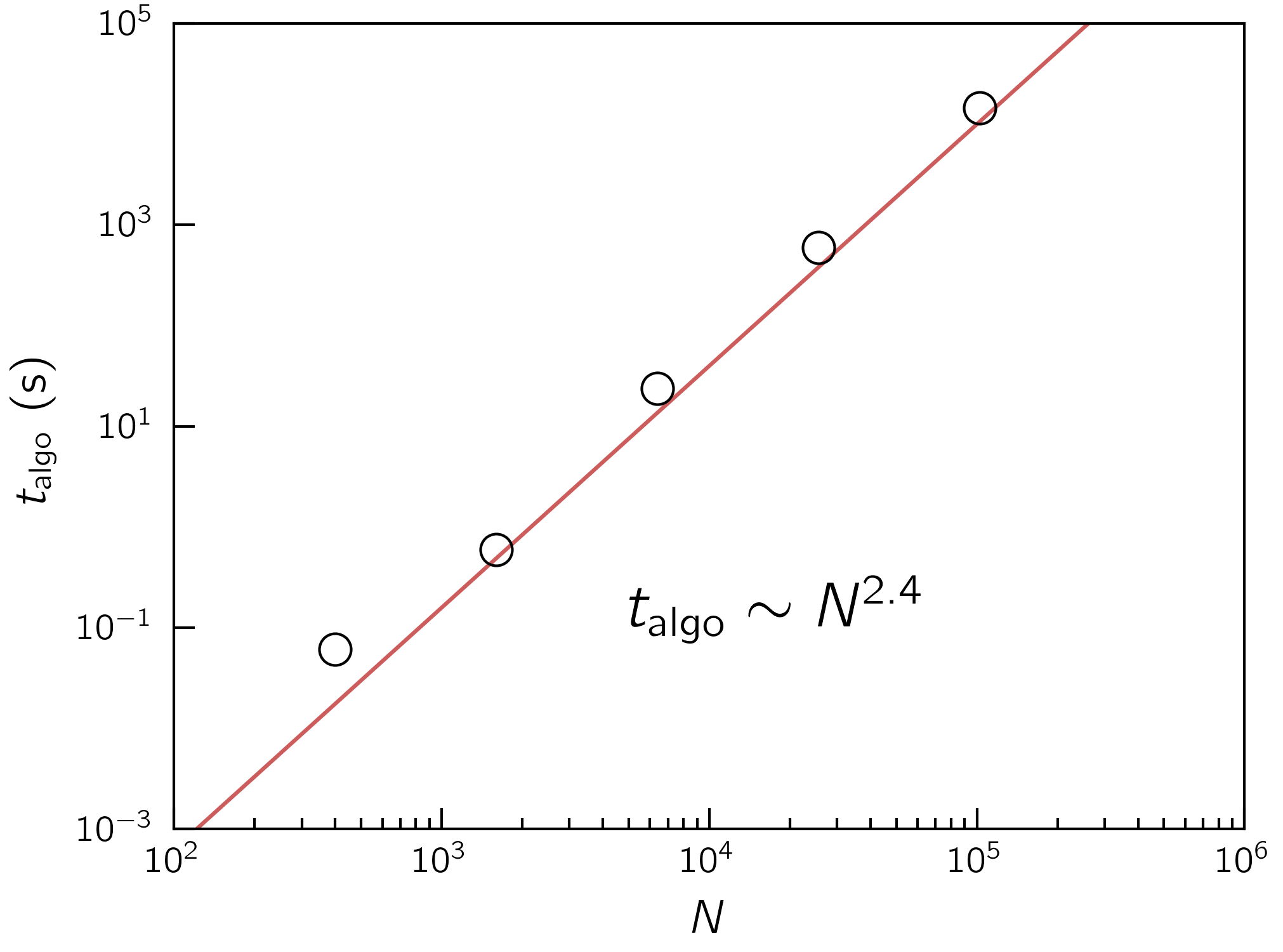}
  \caption{Average time $t_{\rm algo}$ to build a catalog of QLEs in a glass composed of $N$ particles with a block length $\xi\!=\!20$. The solid line indicates the scaling $t_{\rm algo}\!\sim\!N^{2.4}$.}
  \label{fig:complexity}
\end{figure}

We have checked for the complexity of our algorithm by computing the average time $t_{\rm algo}$ to build a catalog in a glass composed of $N$ particles with $\xi=20$, see Fig.~\ref{fig:complexity}. We find that $t_{\rm algo}$ scales with $N^{2.4}$. This result is explained by the combination of the trivial linear scaling of the number of modes in the catalog with a fixed $\xi$ and the additional linear scaling due to the complexity of solving linear equations for each mode. Finally, the "halo" effect is slightly pronounced in larger systems with the more likely presence of very low-frequency modes and the scaling $\xi_{H}\sim1/\omega_\pi$. As a result, one needs more often to repeat step 4 of our algorithm.

\section{Nonphononic spectrum of 3D structural glasses}
\label{ap:3dvdos}

\begin{figure}[h!]
  \includegraphics[width = 0.5\textwidth]{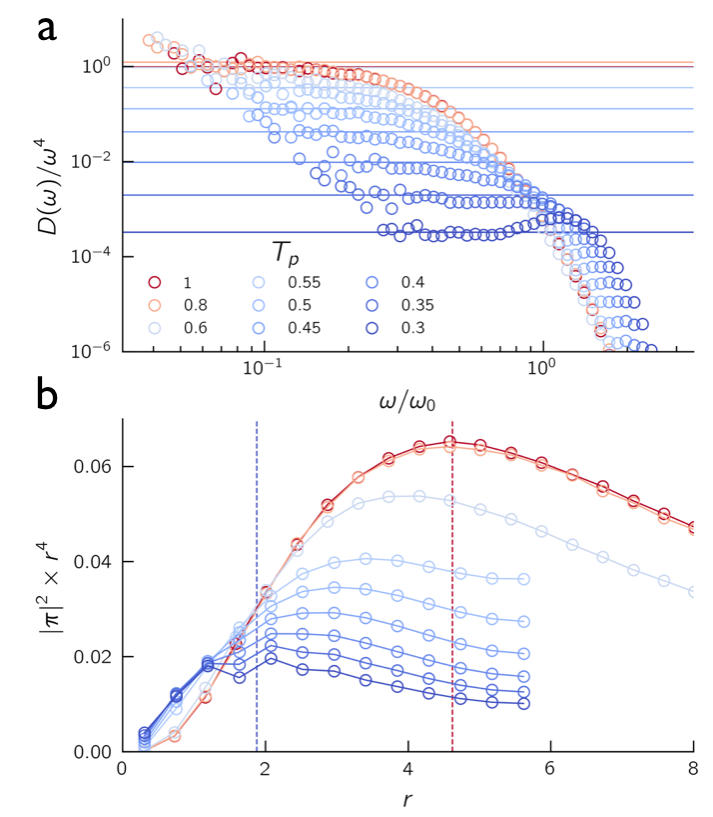}
  \caption{(a) Rescaled PHM vDOS for various parent temperature $T_{\rm p}$ with $\xi=5$. The horizontal lines indicate $A_{\rm g}$. (b)  Rescaled square of the mode amplitude decay $|\pi|^2$ by $r^4$. Vertical dashed lines indicate the reduced core size $\xi_\pi$ when decreasing $T_{\rm p}$.}
  \label{fig:3dvdos}
\end{figure}

Here, we provide additional data for 3D glasses used to extract the nonphononic prefactor $A_{\rm g}$, the mode frequency $\omega_\pi$ and the core size $\xi_\pi$. In Fig.~\ref{fig:3dvdos}(a), we plot the rescaled PHM vDOS to higlight the range of $A_{\rm g}$ (low frequency plateau) as a function of the parent temperature $T_{\rm p}$. Consistent, with Ref.~\cite{rainone2020pinching}, we observe a drop of 3 orders of magnitude in $A_{\rm g}$ from hyper quenched glasses to very stable glasses obtained via swap Monte Carlo. 

Extracting the lowest mode of each glassy sample, we compute the the average of the square of the mode amplitude decay $|\pi|^2$. In Fig.~\ref{fig:3dvdos}(b), we plot the rescaled mode decay by its $r^{-4}$ asymptotic scaling. The peak in $r^{4}|\pi|^2$ allows us to extract an estimate for $\xi_\pi$. We find a factor two decrease of the core size of soft glassy of excitations, consistent with the work done in Ref.~\cite{rainone2020pinching} and with the factor two increase of the typical mode frequency $\omega_\pi$.


%

\end{document}